\documentclass[amsmath,amssymb,aps,nofootinbib,superscriptaddress,onecolumn, showkeys]{revtex4-1}

\usepackage{graphicx}
\usepackage{natbib}

\usepackage{xcolor}

\usepackage[utf8]{inputenc}
\usepackage{graphicx}
\graphicspath{ {figures/} }
\usepackage{array}

\newcommand{\degree}{^{\circ}}
\newcommand{\thm}{\bar{\theta}}
\newcommand{\dgamma}{\Delta\gamma}
\newcommand{\cmc}{c^\ast}

\begin{document}


\title{Delayed coalescence of surfactant containing sessile droplets} 


\author{M. A. Bruning}
\email[]{m.a.bruning@utwente.nl}
\affiliation{Physics of Fluids Group, Faculty of Science and Technology, Mesa+ Institute, University of Twente, 7500 AE Enschede, The Netherlands}

\author{M. Costalonga}
\affiliation{Physics of Fluids Group, Faculty of Science and Technology, Mesa+ Institute, University of Twente, 7500 AE Enschede, The Netherlands}

\author{S. Karpitschka}
\affiliation{Max Planck Institute for Dynamics and Self-Organization (MPIDS), 37077 G{\"o}ttingen, Germany}

\author{J. H. Snoeijer}
\affiliation{Physics of Fluids Group, Faculty of Science and Technology, Mesa+ Institute, University of Twente, 7500 AE Enschede, The Netherlands}

\date{\today}




\begin{abstract}

When two sessile drops of the same liquid touch, they merge into one drop, driven by capillarity. However, the coalescence can be delayed, or even completely stalled for a substantial period of time, when the two drops have different surface tensions, despite being perfectly miscible. A temporary state of non-coalescence arises, during which the drops move on their substrate, only connected by a thin neck between them. Existing literature covers pure liquids and mixtures with low surface activities. In this paper, we focus on the case of large surface activities, using aqueous surfactant solutions with varying concentrations. It is shown that the coalescence behavior can be classified into three regimes that occur for different surface tensions and contact angles of the droplets at initial contact. However, not all phenomenology can be predicted from surface tension contrast or contact angles alone,  but strongly depends on the surfactant concentrations as well. This reveals that the merging process is not solely governed by hydrodynamics and geometry, but also depends on the molecular physics of surface adsorption.

\end{abstract}

\maketitle




\section{Introduction}

The coalescence and interaction of sessile droplets has been the subject of intensive study in scientific literature~\cite{Hernandez-Sanchez2012, Eddi2013, Eddi17, Eddi18, Eddi19, Sui, Cira2015, Karpitschka2012, Karpitschka2014}, not least motivated by its relevance for industrial processes such as ink-jet printing or surface processing. These studies have revealed intriguing and frequently counterintuitive physics behind this ubiquitous process.
When two droplets of the same liquid touch, a liquid bridge forms in between them. Capillary forces promote filling  of this bridge and the droplets merge into a single liquid body.
In contrast, sessile drops of two different but miscible liquids do not always merge instantaneously after contact~\cite{Karpitschka2012, Karpitschka2014, Karpitschka2014a}. Under certain conditions the height of the liquid bridge connecting the two droplets does not grow, but remains very thin as compared to the droplets. Connected by a thin liquid neck, the droplets move over the substrate without merging. 
The droplet with the lower surface tension chases the other droplet, and only significantly later, the drops will merge. 
From a free-energy perspective, this is very surprising, as the total free energy of the two miscible drops is certainly lowered by merging into a single droplet.

The mechanism behind this ``delayed coalescence'' has been identified as a competition between capillary and Marangoni flows~\cite{Karpitschka2012,Karpitschka2014}. The difference in surface tension between the two drops causes a Marangoni flow which drains the liquid bridge between the drops. If this flow is strong enough, it compensates the capillary flows that fill this neck. It also causes a motion of the neck toward the droplet with the higher surface tension. Studies have addressed the coalescence behavior of drops of pure organic liquids on completly~\cite{Karpitschka2014} and partially wetted substrates~\cite{Karpitschka2014a}, and of aqueous solutions of liquids with low surface activity~\cite{Karpitschka2010,Karpitschka2012,Cira2015}. It was found that the coalescence behavior is determined by two key parameters, the surface tension contrast between the droplets, and their (mean) contact angles in the moment when they initially touch. Experiments on non-surface active, non-volatile liquids have revealed a single sharp boundary between the regimes of immediate coalescence and delayed coalescence, where delayed coalescence was observed for small contact angles and/or large surface tension differences. Other control parameters like the droplet volume did not show any significant influence on the coalescence behavior~\cite{Karpitschka2010,Karpitschka2012,Karpitschka2014,Karpitschka2014a,Cira2015}.

In this paper we investigate the coalescence behavior of spreading drops of surfactant solutions. This is relevant e.g. in ink-jet printing processes: typical inks are highly complex formulations that almost always contain surfactants. Precisely controlling the coalescence of printed droplets is of key importance for the quality of the final print. It is well known that the physico-chemical behavior of surfactant solutions can be significantly different from ``simple'' mixtures of liquids with low surface activity. The surface tension of surfactant solutions is a highly nonlinear function of concentration, and the kinetics of adsorption and desorption to interfaces play an important role for dynamical processes~\cite{Afsar-Siddiqui2003}. One pronounced feature of spreading surfactant solutions, which is of relevance here, is the frequently observed fingering instability of spreading fronts. This instability develops when drops of surfactant solutions are placed on thin films of water~\cite{Marmur1981}, or when they spread in a high humidity environment~\cite{Cachile_Langmuir}. Instead of the regular spreading behavior with a smooth circular spreading front, fingers will grow out of the front, leading to irregular patterns and complex dynamics~\cite{Troian1989a, Troian1990, Hamraoui2004, Afsar-Siddiqui2003, Cachile_Langmuir, Cachile1999}. Therefore, also a different coalescence behavior could be anticipated.

Here we present a systematic experimental study of the coalescence behavior of sessile drops of surfactant solutions, showing how the coalescence process is altered in the presence of surfactant induced Marangoni flows. We discover three distinct modes on non-coalescence and a dependence on surfactant concentration that cannot be described by the surface tension difference alone. 

The paper is organised in the following way. In section 2 we describe the experimental methods by which we studied the coalescence behavior of surfactant drops. In section 3a we report our observations of different coalescence regimes, section 3b focuses on the motion of the drops in the delayed coalescence state, and in section 3c we report on the duration of this state. In section 4 we discuss and conclude our results.

%
%

\section{Experimental methods}

\begin{figure*}[tb]
\begin{center}
  \includegraphics{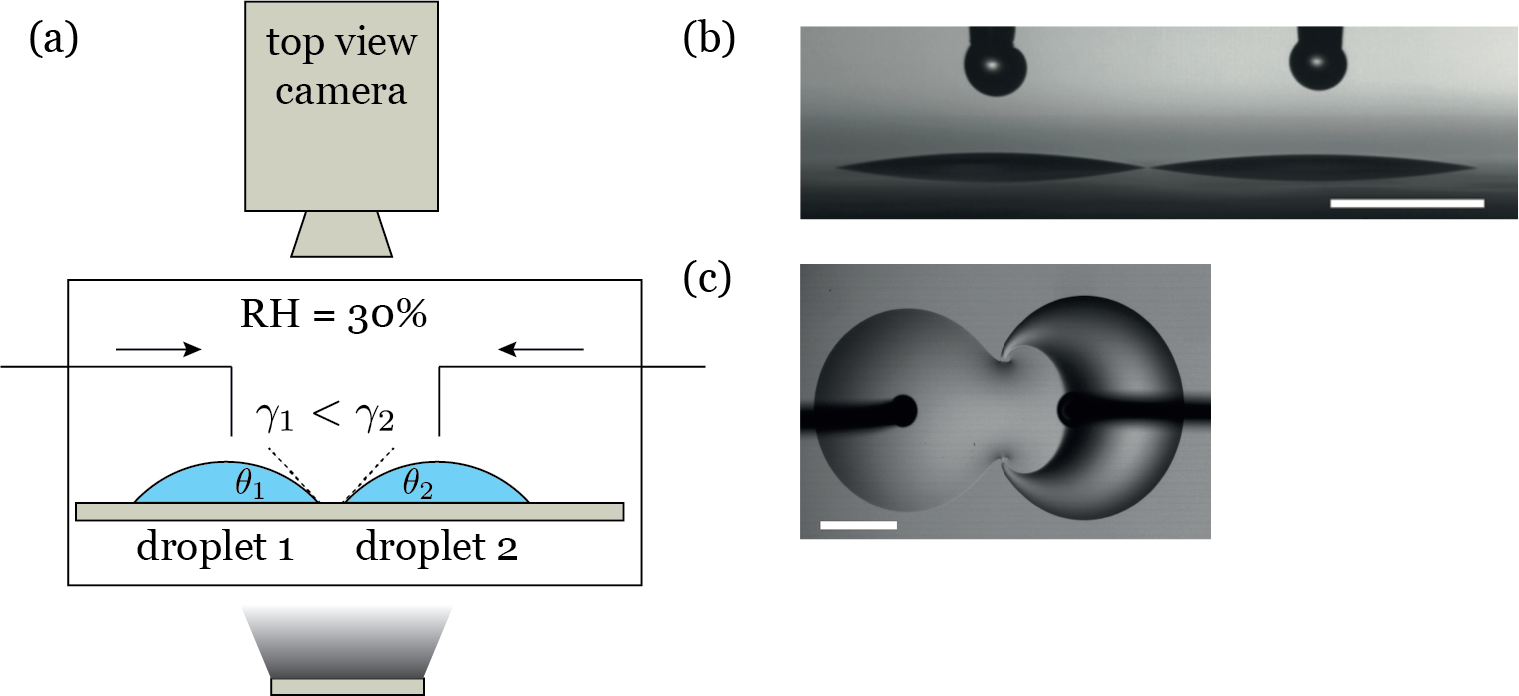}
\end{center}
\vspace{-6mm} \caption{(a) Cross-sectional sketch of the experimental set-up, defining the parameters $\theta_i$ and $\gamma_i$. (b) and (c) show typical images obtained from side and top view cameras, respectively. In both images the scale bar represents 3 mm. The left droplet is pure water, the right droplet contains SDS. Note that in the top view the needle visible in the image is no longer in contact with the droplets. \label{fig: Set-up} }.
\end{figure*}

\begin{figure*}[b]
\begin{center}
  \includegraphics{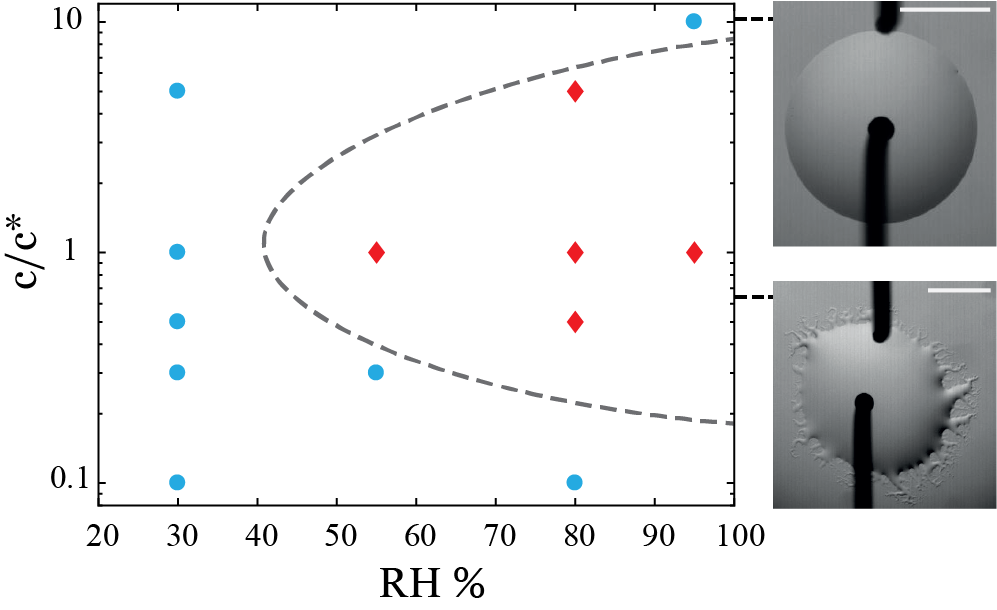}
\end{center}
\vspace{-6mm} \caption{The spreading behavior for a single SDS droplet, depending on relative humdity and surfactant concentration (normalized with the critical micelle concentration $\cmc$). Blue circles represent the stable circular spreading, see upper image (RH = 55$\%$ and 0.3$\cmc$) . Red diamonds depict the fingering case, shown in the bottom image (RH = 55$\%$ and 1$\cmc$). Both images are taken 1.5 s after droplet deposition. The scale bar represents 5 mm. The dashed line between the regimes is intended as a guide to the eye. \label{fig: appendix_RH} }
\end{figure*}

The principle of the experimental procedure is to simultaneously deposit two droplets of surfactant solutions of various concentrations at some initial distance onto a substrate (Fig. \ref{fig: Set-up}). The droplets spread over the substrate until they eventually touch. The (dynamic) solid-liquid-vapor contact angle in that moment can therefore be controlled by adjusting the volume and the initial distance between the drops. As liquids we used solutions of Sodium Dodecyl Sulfate solutions (SDS, Sigma Aldrich, purity 98\% p.a., used as supplied) in MilliQ water (MilliPore Q-Pod, resistivity $18$~M$\Omega$cm) at various concentrations ranging from $0.1\cmc$ to $10\cmc$, where $\cmc = 8.1$~mMol/l is the critical micellar concentration at which the surface tension of the solution saturates~\cite{Mysels1986}. The surface tensions of the solutions are measured by the pendant drop method; analysis is performed with an open-source code~\cite{Berry2015}.
Our experimental set-up is depicted on Fig.~\ref{fig: Set-up}a. Two 5 $\mu L$ pendant droplets of SDS solutions with surface tensions $\gamma_1$ and $\gamma_2$ ($\gamma_1 < \gamma_2$) are created at tips of $0.9$~mm-diameter needles using a volume-controlled syringe pump.  To ensure a smooth deposition on the substrate, the substrate is lifted by a motorized stage to collect the droplets. As substrates we used standard glass microscope slides (Menzel-Gl\"aser, 26 x 76 mm) which were cleaned by immersion in piranha solution for 30 minutes, rinsing and sonicating them thoroughly with MilliQ water afterwards. This consistently yielded completely wetted surfaces on which the spreading of pure water droplets precisely followed Tanner's law.

The spreading and coalescence behavior was monitored by simultaneously recording side and top view with digital cameras that were triggered synchronously at 150 frames/s with a pulse generator (BNC model 575). 
The contact angles were measured optically using the side-view recordings (PCO-1200s camera, Zeiss 60 mm macro lens) as shown in Fig.~\ref{fig: Set-up}b. An example of top-view recordings (Ximea XiQ MQ013MG-ON camera, Sigma 50 mm macro lens) is shown in Fig.~\ref{fig: Set-up}c, featuring a typical case of delayed coalescence at $t = 0.2$ s after initial contact.

Controlling the ambient relative humidity, $RH$, is critical to our experiments. Depending on $RH$, fingering instabilities, similar to those reported e.g. by Marmur and Lelah~\cite{Marmur1981}, can arise during the spreading of our SDS solutions. Consequently, all experiments are performed in a closed chamber, where relative humidity is maintained at a constant value by a home built apparatus, using a feedback loop to control the $RH$. The $RH$ is constantly monitored with a sensor inside the chamber. When the measured $RH$ is lower than the target value, humid air (generated by bubbling through water in a gas wash bottle) is gently injected into the chamber.  Conversely, dry nitrogen gas is injected to lower $RH$. 
In Fig. \ref{fig: appendix_RH} we report our observations regarding the occurence of fingering instabilities with respect to $c$ and $RH$ for the spreading of single SDS solution droplets on hydrophilic glass slides. Fingering was not observed at very low or very high surfactant concentrations. It appears for intermediate concentrations at humidities larger than $40\%$. The range of concentration for which fingering was observed increases with relative humidity, as has also been reported by Cachile and Cazabat~\cite{Cachile1999a_colloids}. For low ambient humidities $RH = 30$ \% and concentrations ranging from $0.1\cmc$ to $5\cmc$, we observe stable spreading that follows Tanner's law. Therefore, all coalescence experiments that we will present in the following were performed at $RH = 30$~\% in order to avoid this instability and initiate the coalescence process with well-defined drop geometries.




\section{Results and interpretation}

\subsection{Coalescence regimes}

\begin{figure*}[t]
\begin{center}
  \includegraphics{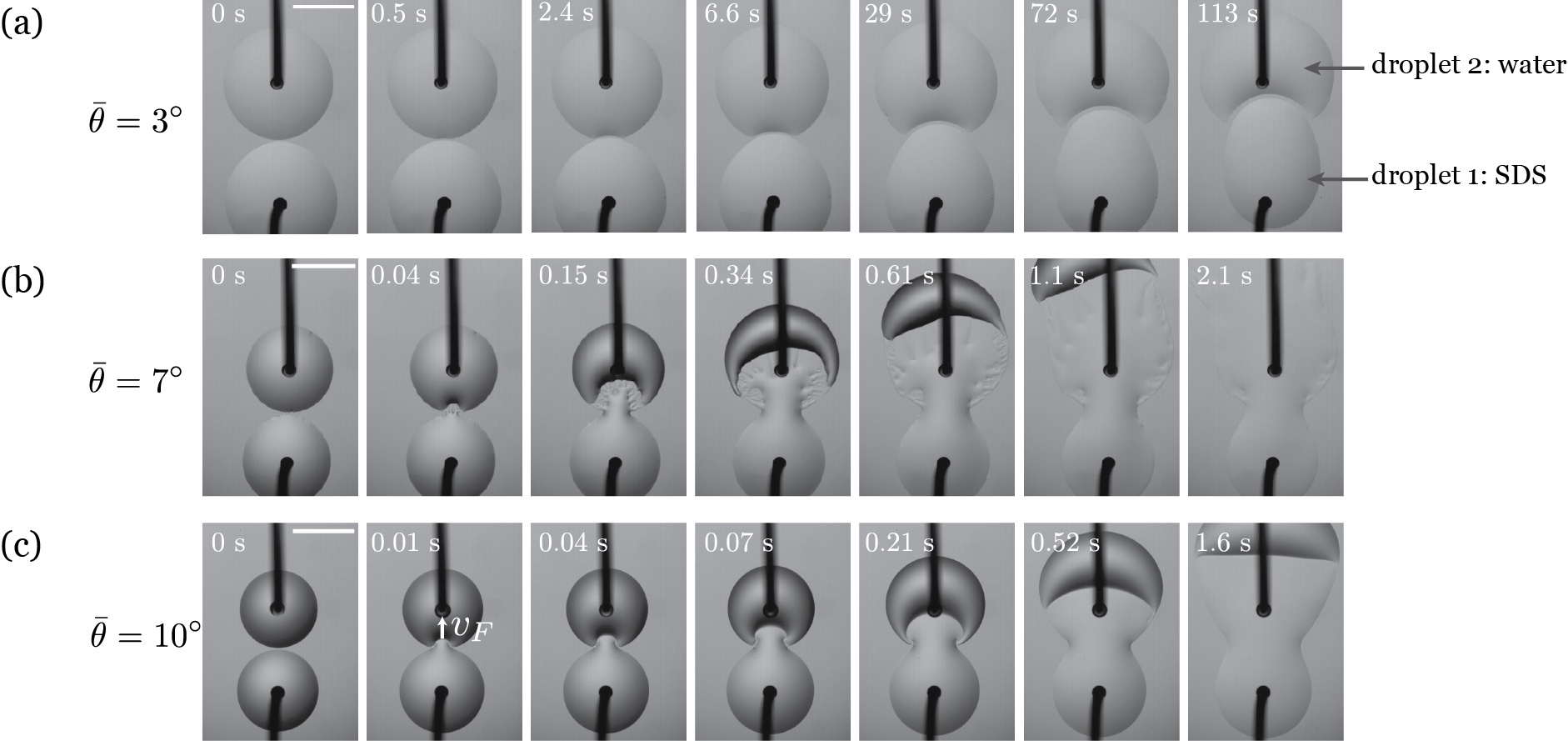}
\end{center}
\vspace{-6mm} \caption{Coalescence behavior for a surfactant drop (droplet 1) and a pure water drop (droplet 2), $\gamma_2>\gamma_1$. Three regimes are observed for different contact angles: a) precursor-mediated interactions, b) fingering regime and c) stable delayed coalescence. $t$=0 refers to the moment the droplets touch. In each series the upper drop consists of pure water and the bottom drop is the surfactant solution, in all cases $\dgamma$ = 13 mN/m. In all series the scale bar represents 3 mm. 
\label{fig: timelapses phase} }
\end{figure*}

There are three main control parameters in our experiments which control the coalescence behavior. These are the mean contact angle at which the two droplets meet, $\thm = (\theta_1+\theta_2)/2$, their surface tension difference $\dgamma = \gamma_2-\gamma_1$, and the surfactant concentrations $c_1$ and $c_2$. 
In all experiments that we performed, a state of delayed coalescence is observed, provided that the two drops are not from the same solution (i.e., $\dgamma \neq 0$), even for the smallest tested $\dgamma \sim 2$~mN/m. In all cases with $\dgamma \neq 0$, we observed three different modes of delayed coalescence. In contrast to previous observations with non-surface active miscible liquids~\cite{Karpitschka2012,Karpitschka2014,Karpitschka2014a}, different behaviors could be observed depending on the surfactant concentrations, even with identical $\thm$ and $\dgamma$.
Therefore we split the coalescence experiments into two main groups: (i) experiments where only one droplet initially contained surfactant, while the other one was pure water, and (ii) those where both droplets were prepared from surfactant solutions.

We begin with a description of case (i), the coalescence behavior of a SDS containing droplet (drop 1) with a pure water droplet (drop 2), i.e. initially $c_2=0$ and $\gamma_2 = 72$~mN/m. Thus the surfactant concentration $c_1<\cmc$ of drop 1 and the surface tension difference between the drops are not independent, and we will representatively use $\dgamma$ in the following. Depending on $\dgamma$ and $\thm$, three distinct regimes are observed. A time series of top view images for each regime is shown in Fig.~\ref{fig: timelapses phase}. See Supplemental Material at [...] for a movie of each regime.

At low contact angles (Fig.~\ref{fig: timelapses phase}a, $\thm = 3\degree$), the lower surface tension drop (drop 1, containing SDS) is attracted toward the higher surface tension drop (pure water, drop 2), while the water drop is pushed away from the SDS drop. However, an apparent gap separates the two drops, and the macroscopic liquid bodies remain separated. In this configuration, the droplets travel together on the substrate without coalescing. For volatile binary mixtures, interaction at a distance had previously been attributed to Marangoni flows induced by evaporation and condensation and termed vapor-mediated interaction~\cite{Bangham1938,Cira2015}. This could explain the response of the surfactant droplet because water vapor is enriched in the interstitial region. However, the response of the pure water droplet can not result from water evaporation or condensation alone. Water vapor does not affect the surface tension of pure water, and volumetric or thermal consequences of evaporation or condensation have a much weaker impact on the spreading of a one-component liquid~\cite{Bonn2009} than what is observed here. On the other hand, SDS is essentially non-volatile. This suggests the importance of a wetting precursor in the transport phenomena between the two droplets. The presence of the precursor around the spreading surfactant is also described in~\cite{Cachile_Langmuir,Frank1995}.

Increasing the mean contact angle $\thm$ reduces and finally removes this gap between the droplets, so that they spread into macroscopic contact. Fig.~\ref{fig: timelapses phase}b shows an experiment at $\thm = 7\degree $. A fingering instability occurs in the connecting region, which is reminiscent of the fingering instability observed when a droplet of surfactant solution is deposited onto a film of water \cite{Marmur1981, Troian1989a}. At the same time, the water drop is being propelled away from the SDS drop, frequently traveling distances that exceed the footprint size of the two droplets. It is interesting to note also that the dynamics of this regime is quite different from the previous case at lower $\thm$: the water droplet moves $\sim 100$ times faster as compared to the precursor-/vapor-mediated interaction.

Finally, at even higher contact angles, the fingering instability disappears, and a smooth, stable, and  moving neck separates the droplet, similar to delayed coalescence for non-surface active liquids. Fig.~\ref{fig: timelapses phase}c shows an experiment at $\thm = 10\degree$, in which the front between the two droplets remains smooth. The water droplet advances, set into motion by the Marangoni flow due to the lower surface tension of the SDS-droplet. At the same time, the SDS droplet flattens into a film. The overall process lasts for several tens of seconds.

\begin{figure*}
\begin{center}
  \includegraphics{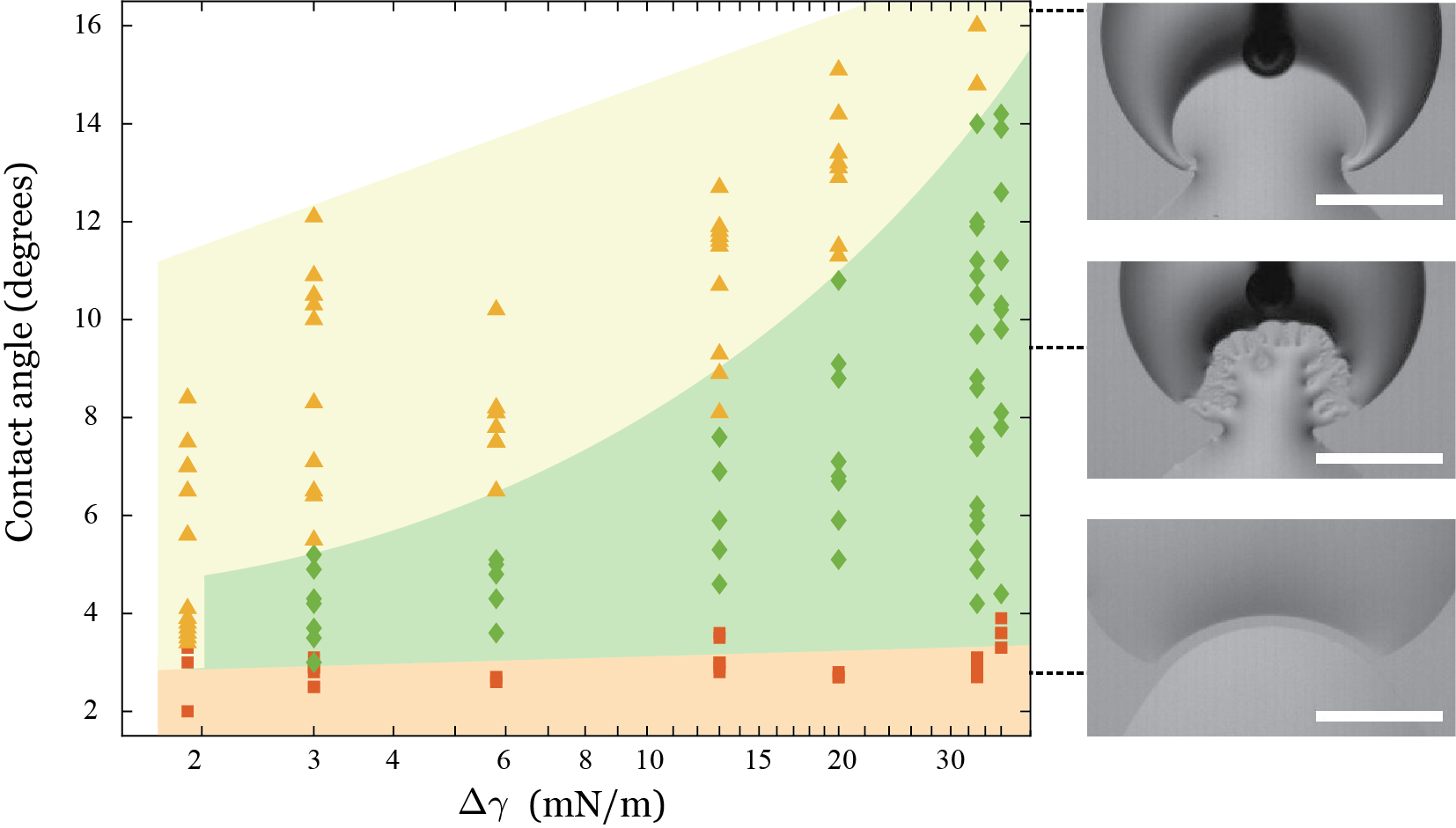}
\end{center}
\vspace{-6mm} \caption{Phase diagram of the coalescence behavior of a water drop and a SDS-drop. Both the surface tension difference $\Delta\gamma$ and contact angle are varied. Red squares: vapor-/precursor mediated interaction, green diamonds: fingering instability, yellow triangles: stable delayed coalescence. The color regions are intended as a guide to the eye. The insets show zooms of the top view of the contact region between the two drops (scale bars: 3 mm).  \label{fig: water-SDS} }
\end{figure*}

We construct a phase diagram of these regimes in Fig.~\ref{fig: water-SDS}, by repeating the coalescence experiments for various $\dgamma$ (adjusted by the SDS concentrations in the SDS drop) and $\thm$. It appears that vapor-/precursor-mediated interactions occur only for contact angles below $3\degree$, regardless the surface tension difference between the droplets (red squares). It occurs already at the lowest concentration $c_2 = 0.05$~mMol/l we used, but was absent for two pure water droplets. At higher contact angles, there is no gap between the drops, but fingers appear, and the motion of the water droplet is much faster (green diamonds). These changes clearly appear within a fraction of a degree of $\thm$, rendering a sharp transition from vapor-/precursor-mediated interaction to the fingering regime. Increasing the contact angle further, fingering eventually disappears and a smooth front between the drops appears. The transition to this mode is not sharp: as $\thm$ increases, fingers are first removed from the center of the connecting front, but still appear on its sides, until they gradually disappear completely. Yellow triangles indicate experiments for which no fingering in the centre was visible anymore. This transition depends on $\dgamma$, and fingering occurs on a wider range of contact angles as $\dgamma$ increases. This trend is consistent with previous observations 
for surfactant droplets spreading on water films. There, higher surface tension differences were required to observe the instability on thicker films \cite{Afsar-Siddiqui2003}. In the present case, the contact angle adopts a role reminiscent of the film thickness.

\begin{figure}[t]
\begin{center}
  \includegraphics{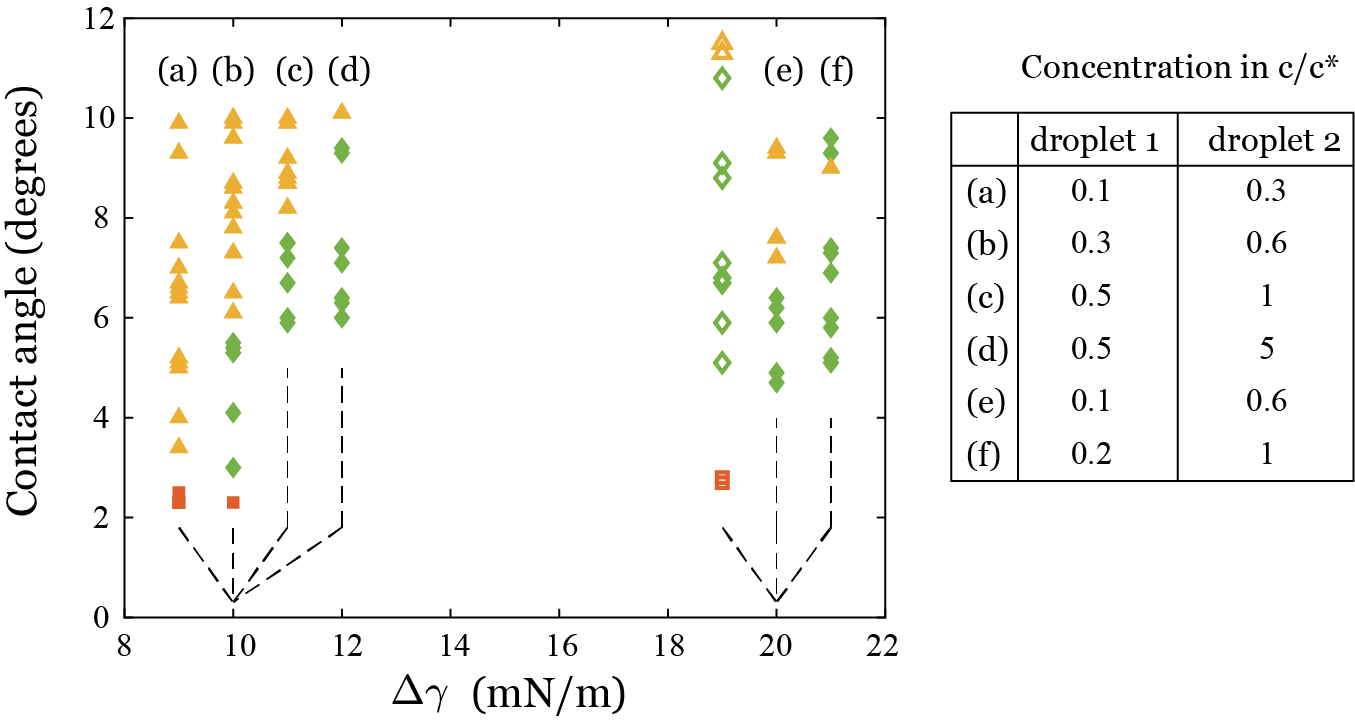}
\end{center}
\vspace{-6mm} \caption{Phase diagram of two coalescing droplets of different SDS concentrations, the same legend as figure \ref{fig: water-SDS} applies for the regimes. Several concentration couples are used to obtain series at $\Delta\gamma$ = 10 mN/m and 20 mN/m. From left to right we move to higher concentrations. The table specifies the concentrations, in terms of the critcal micelle concentration $\cmc$. The open symbols at 20 mN/m represent the water-SDS case from figure \ref{fig: water-SDS}, included for comparison. \label{fig: SDS-SDS} }
\end{figure}

Next we consider case (ii), the coalescence behavior of two droplets of surfactant solutions. Interestingly, if SDS is initially present also in the second droplet, the boundary between fingering and stable delayed coalescence is strongly altered, whereas vapor-/precursor-mediated chasing appears as before. Fig.~\ref{fig: SDS-SDS} shows how this affects the phase diagram. We prepared droplet pairs with two values of $\dgamma = \{10,20\}$~mN/m, using several different surfactant concentrations to realize each $\dgamma$. The corresponding results for case (i) and $\dgamma = 20$~mN/m are indicated in this figure as well for comparison.

From Fig.~\ref{fig: SDS-SDS} it is clearly visible that the surfactant concentration has a strong impact on the coalescence behavior, even at identical $\dgamma$.
A slight amount of surfactant in the previously pure water droplet reduces the range of the fingering instability with respect to $\thm$. For $\dgamma = 10$ mN/m, it is even possible to completely avoid the fingering destabilisation, in contrast to case (i) where it was always observed except for $\dgamma = 2$~mN/m. However, increasing the mean surfactant concentration further, expands the range of the fingering instability. Also, increasing $c$ beyond $\cmc$ for one droplet (series (c) and (d) at $\dgamma = 10$~mN/m in Fig.~\ref{fig: SDS-SDS}), yet keeping the same $c$ in the other droplet, expands the fingering regime. These effects show that, for surfactant solutions, the physics on a molecular scale have a direct influence on the macroscopic hydrodynamics of the coalescence process.

\subsection{Front velocity}

Next we focus on the characteristics of the motion of the neck between droplets during stable delayed coalescence (yellow region in Fig.~\ref{fig: water-SDS}).
As the two droplets meet, the higher surface tension droplet (droplet 2) ``pulls'' on the lower surface tension drop (droplet 1), with a clearly distinguishable front (the ``neck'') separating them. After the neck has formed, it travels with a certain velocity $v_F$ (see Fig.~\ref{fig: timelapses phase}a). This velocity is not constant over time. It reaches its maximum value almost instantaneously (below the temporal resolution of our imaging system) after initial contact and continuously decays thereafter. In the following we focus here on the initial (maximum) velocities.

\begin{figure*}
\begin{center}
  \includegraphics{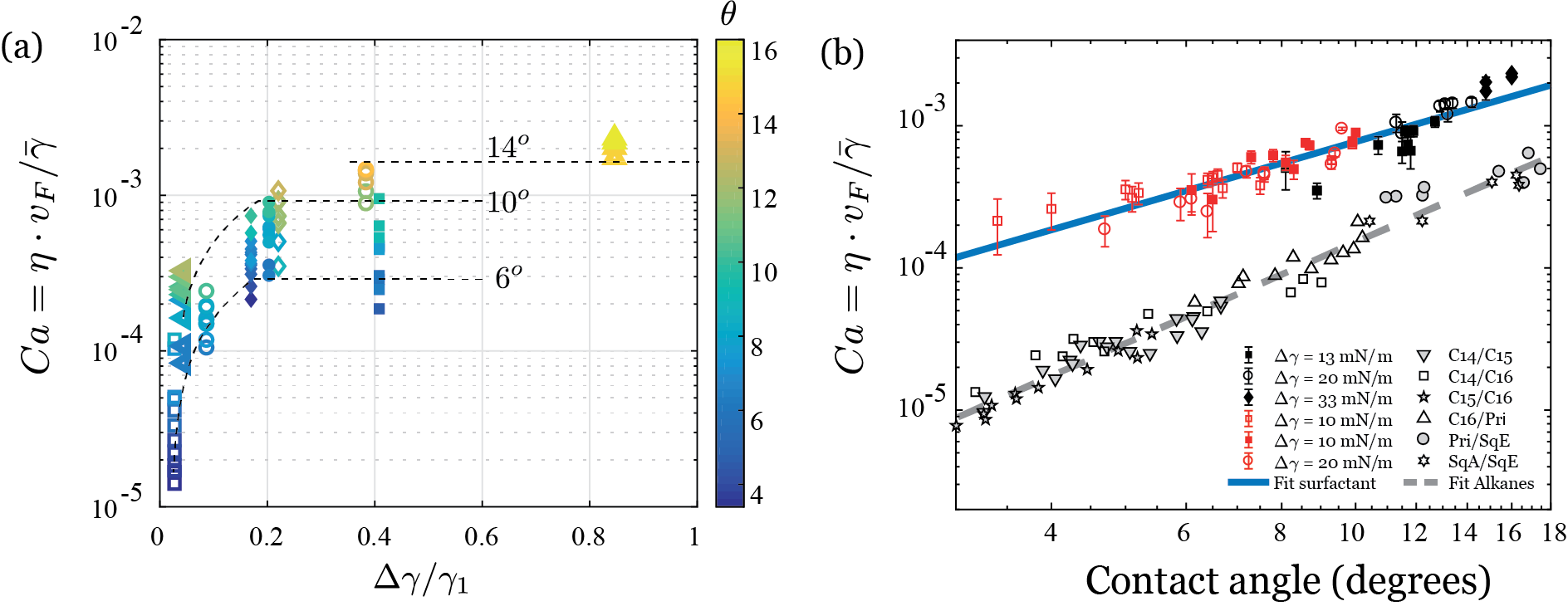}
\end{center}
\vspace{-6mm} \caption{The capillary number of the front of the moving drop (droplet 2) versus reduced surface tension difference $\Delta\gamma/\gamma_1$ ($\gamma_1<\gamma_2$) and contact angle, for the delayed coalescence regime. In a) the open symbols represent the coalescence of a water and SDS-droplet, closed symbols represent two SDS-droplets. The colorbar shows the contact angle in degrees and the dotted lines serve as a guide for the eye to see the trend for $Ca$ for $\theta$ = 6, 10 and 14$^\circ$. In b) the black series are water-SDS and the red series are the SDS-SDS case, inluding the best fit as a blue line. The second column of the legend shows the results for delayed coalescence of different alkane pairs from ~\cite{Karpitschka2014}, inserted for comparison. \label{fig: Ca-figures} }
\end{figure*}

We non-dimensionalize $v_F$ with the capillary velocity and obtain the capillary number $Ca = \eta v_F/\bar{\gamma}$, with $\bar{\gamma} = (\gamma_1+\gamma_2)/2$. Fig.~\ref{fig: Ca-figures}a shows that $Ca$ initialy increases with $\dgamma$, but saturates at $\dgamma/\gamma_1\sim 0.2$, for all mean contact angles $\thm$ we explored ($4\degree \leq \thm \leq 16\degree$). This is consistent with previous observations on non-surface active liquids~\cite{Karpitschka2012}, although in the SDS case, higher $\dgamma$ are required for velocity saturation. Also consistent with~\cite{Karpitschka2012}, experiments with pure water vs. SDS solution (case (i), open symbols on Fig.~\ref{fig: Ca-figures}a), and those with two surfactant-containing droplets (case (ii), filled symbols on Fig.~\ref{fig: Ca-figures}a) yield identical $v_F$, provided that the surface tension difference remains the same. This indicates that the front velocity is determined only by $\thm$ and $\dgamma$, and not by the molecular interactions of the surfactant molecules.

On Fig.~\ref{fig: Ca-figures}b we plot the saturated values of $Ca$ as a function of $\thm$. In this regime, all data, irrespective of $\dgamma$ and $c$, collapses onto a single master curve. This is similar to the case of miscible liquids of low surface activities reported previously~\cite{Karpitschka2012} and suggests that the driving mechanism for the motion of the neck is the Marangoni flow. However, the values of $Ca$ for the case of surfactant solutions studied here differs from those of previously reported experiments on simple non-polar liquids~\cite{Karpitschka2014}. The experimental results for various combinations of room-temperature liquid alkanes (see~\cite{Karpitschka2014} for a detailed description of these experiments) are included here as well. Both datasets show a saturation of $Ca$ with respect to $\dgamma$, but the saturated values of $Ca$ are significantly different. For small $\thm$, the front between droplets of different alkanes moves by about an order of magnitude slower than for aqueous SDS solutions. Also the trend of $Ca$ with respect to $\thm$ is different for both cases: aqueous surfactant solutions are best fitted by  $Ca \sim\thm^{1.6}$, the alkane data by $Ca \sim\thm^{2.4}$. The main difference between both cases are the relevant transport phenomena: in the case of surfactants, surface advection and diffusion presumably becomes important in addition to bulk advection and diffusion. Also, changes in surface tension depend on the ad- and desorption kinetics of the surfactant molecules, which are much slower as compared to the almost instantaneous equilibrium of alkane mixtures. Last but not least, electrostatic interactions could play an important role for an ionic surfactant like SDS. Either of these effects could have an impact on the height profile and the local composition of the liquids, and thus also on the front velocities.

\begin{figure*}
\begin{center}
  \includegraphics{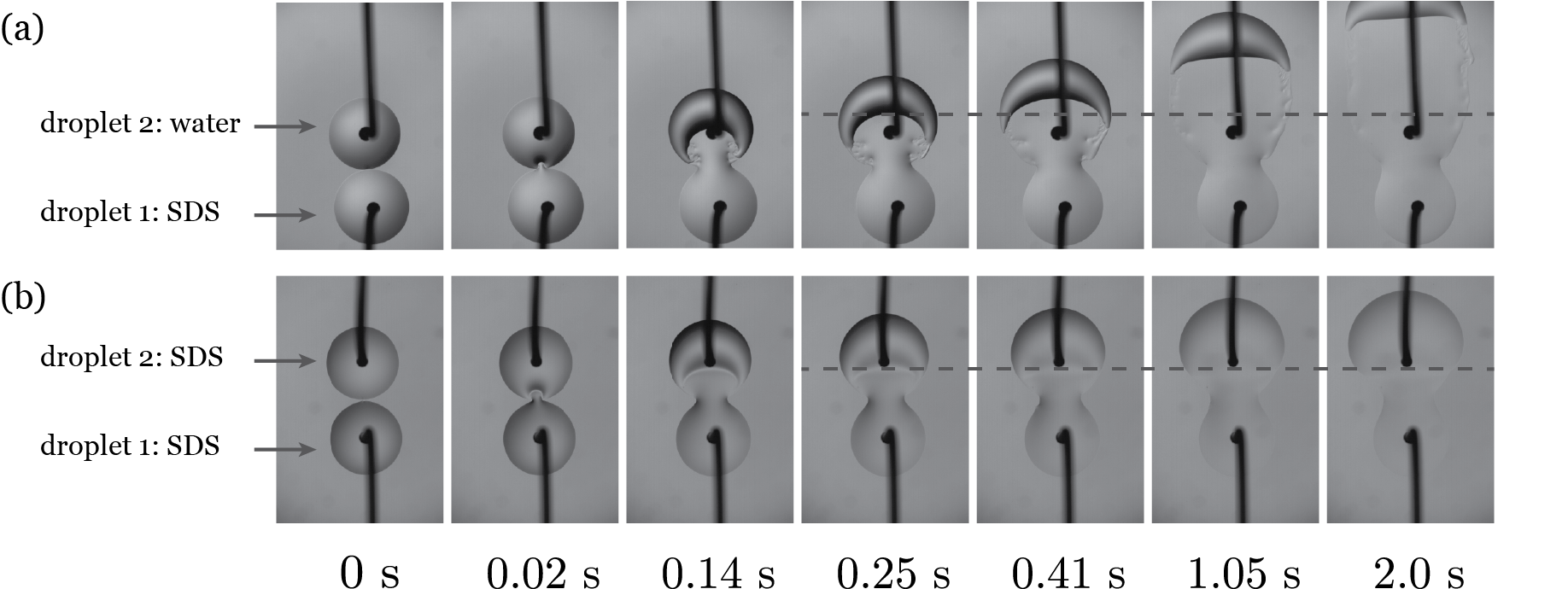}
\end{center}
\vspace{-6mm} \caption{Image sequences for two experiments with approximately the same surface tension difference: a) shows the coalescence of a pure water droplet with a SDS solution droplet (0.3 $\cmc$), with $\Delta\gamma$ = 13 mN/m, whereas b) shows the coalescence between two SDS solutions droplets (droplet 1: 0.6 $\cmc$ and droplet 2: 0.3 $\cmc$), with $\Delta\gamma$ = 10 mN/m. In both cases a) and b), $\thm$ = 8$^\circ$. The time at which the images are taken are labeled below (the same for a and b). The black dashed line indicates the position of the front separating the two droplets, at $t = 0.25$ s.\label{fig:duration}. In case b) the front does not move any more after $t\sim0.1-0.3 $ s.}
\end{figure*}

\subsection{Duration of delayed coalescence state}

In the previous section we have shown that the initial neck velocities during delayed coalescence depend only on $\dgamma$ and $\thm$, but show no explicit dependence on the concentrations. However, the dynamics at later times differ dramatically. 
Fig.~\ref{fig:duration} shows a direct comparison between two experiments with nearly identical $\dgamma$ and $\thm$, but with different SDS concentrations. The case of pure water against SDS ($c_1 = 0.3\cmc$) is shown on panel a, while on panel b, both droplets contain SDS ($c_1 = 0.6\cmc$ and  $c_2 = 0.3\cmc$). While the initial front velocities are comparable for both cases, the lifetime of the delayed coalescence is significantly different. In case (a), pure water against SDS solution, the motion of the neck and water drop away from the SDS drop is sustained for a long time. The SDS drop flattens into a film, covering the trace of the water droplet. The neck typically remains stable until the SDS droplet has drained completely, which may take up to the order of a minute. The water droplet may travel distances that are an order of magnitude larger than the initial droplet radius. Duration and distance increase with the initial $\dgamma$. In case (b) however, with two SDS droplets, the duration of delayed coalescence is much shorter, even though $\dgamma$ and the initial neck velocities are the same as in case (a). At $t\sim0.1-0.3 $ s, the neck motion has stopped and even starts to recede. Then the neck gradually disappears and the drops merge, visible by the decaying contrast between the two drops. In the Supplemental Material at [...] we provide the corresponding movies from Fig.~\ref{fig:duration}.

\begin{figure*}
\begin{center}
  \includegraphics[width=0.97\textwidth]{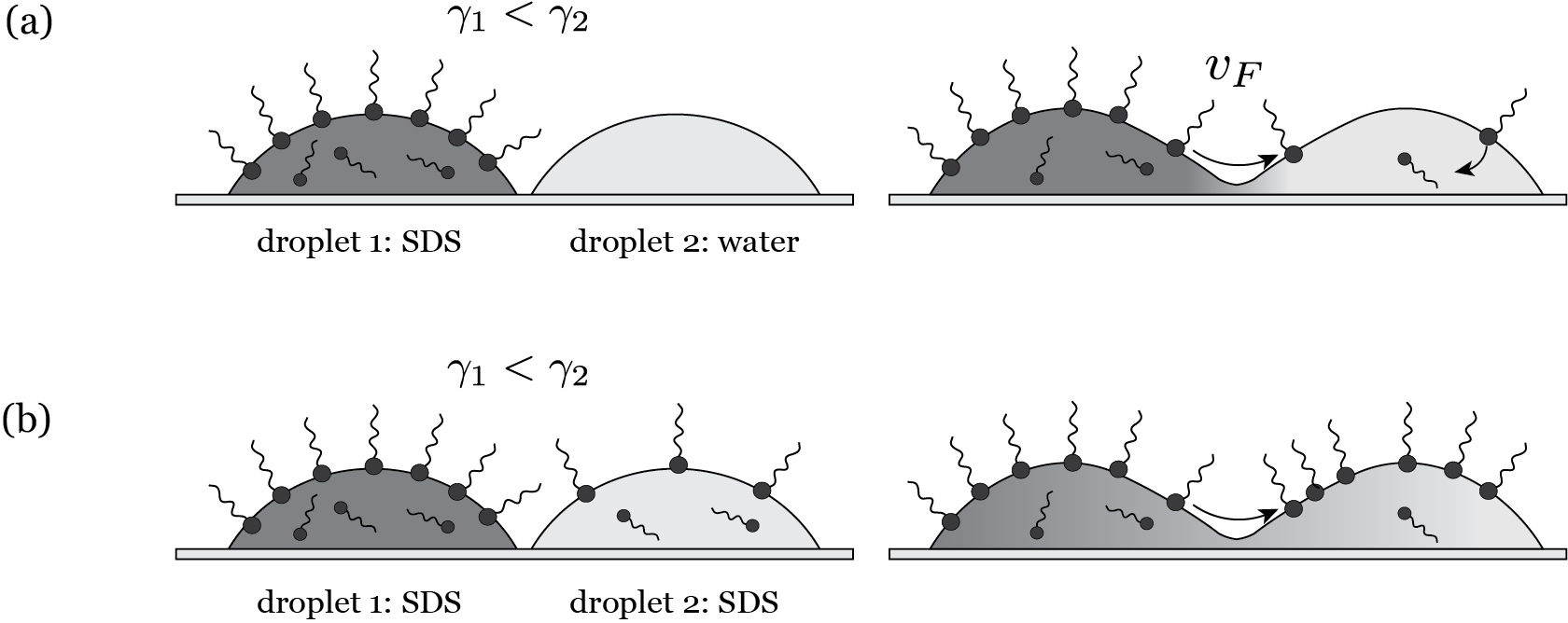}
\end{center}
\vspace{-6mm} \caption{Mechanism of lifetime difference between case (a) and (b) explained. Figures a) and b) show the difference in surfactant coverage. Figure b) shows the accumulation of surfactans at the surface of droplet 2 in a schematic way. In all images size of the surfactant molecules is exaggerated as well as the number of surfactants at the interface as compared to the bulk. }\label{fig:theory}
\end{figure*}

To rationalize the difference in the duration of the delayed coalescence state in the two cases, we consider the surfactant coverage of the front (downstream) droplet, which has the higher surface tension (droplet 2), see Fig.~\ref{fig:theory}. The advection onto an initially surfactant-free surface is maintained for a long time. The surfactant that is advected from droplet~$1$ has an empty surface to cover and an empty bulk liquid to desorb into. This process is schematically depicted in Fig.~\ref{fig:theory}a. In the case of coalescence between two SDS containing droplets, both surfaces are initially covered by surfactant in equilibrium with the respective bulk concentration (see Fig.~\ref{fig:theory}b). Even though the bulk concentration of droplet 2 is rather low in our experiments (at lowest 0.1$\cmc$), the surface coverage is already 2/3 in terms of its saturation value \cite{Sasaki1975}. We propose that in this case, Marangoni convection will rapidly compress the downstream surface and lower its surface tension until coverage reaches saturation. This could be described as a traffic jam-like situation on the surface of droplet 2, since surfactant desorption occurs on much slower timescales~\cite{Chang1995}. Similarly, the upstream surface expands and its surface tension dynamically increases as long as adsorption is comparably slow.
As a consequence, the surface tension contrast would decrease rapidly, and the associated Marangoni flow ceases and the drops merge. A corroboration of this hypothesis would require an independent measurement of the desorption kinetics of SDS where surface and bulk are far from equilibrium. While substantial literature exists on adsorption kinetics~\cite{Chang1995,Casandra2017}, desorption rates of SDS have not been directly measured. The only available data were inferred indirectly by fitting kinetic models to adsorption measurements close to equilibrium. These timescales lie beyond tens of seconds~\cite{Chang1995}, but the correct values for our case might differ to some extent. Nonetheless, with the proposed mechanism of advective surface compression, we can robustly estimate the time required to reach saturation, by comparing the typically observed surface velocities to the droplet diameter. With a front velocity $v_F \sim 50$~mm/s and a droplet footprint radius $R \approx 5$~mm, this gives $\tau = R/v_F = 0.1$ s. Since advection and desorption timescales differ by orders of magnitude, this suggests that the surface reacts mostly elastically. The advective time scale also agrees well with the experimental observation in Fig.~\ref{fig:duration}, where we noted that $\tau $ must be 0.1 - 0.2 s. Furthermore, we also observed a decrease of $\tau $ with contact angle in our experiments . This matches the fact that the front velocity increases with contact angle which would lead to a faster saturation of the surface coverage.

\section{Discussion \& Conclusion}

This experimental study shows that the coalescence of sessile surfactant droplets exhibits a rich phenomenology. Depending on the mean contact angle of the drops and their surface tension difference, we observed 3 different regimes of delayed coalescence: precursor-mediated interaction for droplets with small contact angles, a fingering regime for intermediate contact angles and a strong surface tension contrast between the droplets, and stable delayed coalescence for larger contact angles and weaker surface tension contrast. We also show that the concentration of surfactant alters the lifetime of delayed coalescence.

For small dynamical contact angles, we observed a regime where the droplets interact at a distance, without a visible, macroscopic contact of their liquid bodies. Importantly, this interaction is bi-directional: not only is the SDS-containing droplet attracted toward the water droplet, but also the pure water droplet is repelled by the SDS containing droplet. This excludes a pure vapor mediated process~\cite{Cira2015} since SDS is virtually non-volatile. Most likely, SDS transport is precursor mediated. Accordingly, the macroscopic response of the second contact line offers a novel way to probe the properties of ultra-thin wetting precursors.

We have also investigated the dynamics of a moving neck in the situation of stable direct drop-drop contact. The velocity of the moving neck saturates at $\dgamma/\gamma_1\sim 0.2$, and does not depend on surfactant concentration. This saturation was also observed in delayed coalescence of miscible liquids with low surface activity~\cite{Karpitschka2012}.
However, for the duration of this delayed coalescence state, the concentration of surfactant in the two drops plays an important role. The movement of the neck stops much faster and coalescence proceeds, if both drops initially contain surfactant, as compared to the case where one drop initially consists of pure water. We propose that this direct influence of concentration is due to the advective accumulation of surfactant caused by the Marangoni flow, since the timescale of desorption is much slower than the timescale of advection, and the surface reacts mostly elastically to its compression or expansion. The accumulation reduces the surface tension differences between the two drops, the associated Marangoni flow decreases, and the movement stops. The time scale of advective accumulation until saturation can be estimated and matches well with the experimentally observed lifetime.

In this paper we presented an overview of the basic phenomenology observed in coalescence of sessile surfactant droplets which opens up many possible directions for future investigations. The multi-scale nature and the coupling between bulk and surface terms render it an interesting problem for numerical modelling, to see, for instance, whether a velocity saturation or a coverage saturation would be observed with a minimal model like lubrication theory. Clearly, future experimental work should involve different type of surfactants, to explore the influence of charge and stearic interactions, or to tune adsorption and desorption time scales. Importantly, dynamic surface-compression or subphase-exchange experiments could provide directly measured calibrations of adsorption kinetics.

In conclusion, we have shown how surface active molecules not only alter surface tension, they also change the relevant physical processes at the interface, giving molecular processes an important and directly measurable influence on macroscopic dynamics. There is a coupling between hydrodynamics and physico-chemical processes, such as adsorption/desorption kinetics or electrostatic interactions that give rise to a complicated surface rheology; surface tension depends on the dynamical expansion or contraction of the local surface area, and hydrodynamic boundary conditions can no longer be described by simple linear constitutive equations.

\section{Acknowledgements}

MC and JHS acknowledge financial support from ERC (the European Research Council) Consolidator Grant No. 616918. SK acknowledges financial support from the Max Planck -- University of Twente Center \emph{Complex Fluid Dynamics -- Fluid Dynamics of Complexity}.

\bibliographystyle{unsrt}
\bibliography{Bib_Bruning}

\end{document}